\documentclass{pasj00}
\draft

\begin{document}
\SetRunningHead{Author(s) in page-head}{Running Head}
\Received{2004/08/06}
\Accepted{2005/02/16}

\title{Time Evolution of Relativistic Force-Free Fields Connecting a
Neutron Star and its Disk}

%%% begin:list of authors
\author{Eiji \textsc{ASANO}}
\affil{Graduate School of Science and Technology,
  Chiba University,\\ 1-33 Yayoi-cho, Inage-ku, Chiba 263-8522}
\email{asano@astro.s.chiba-u.ac.jp}

\author{Toshio \textsc{UCHIDA}}
\affil{2-9-1-301, Shikatebukuro, Minami-ku, Saitama 336-0031}\email{uchidat@jcom.home.ne.jp}
\and
\author{Ryoji {\textsc MATSUMOTO}}
\affil{Department of Physics, Faculty of Science, Chiba University,\\
1-33, Yayoi-cho, Inage-ku, Chiba 263-8522}\email{matumoto@astro.s.chiba-u.ac.jp}
%%% end:list of authors

%%% `\KeyWords{}' always has to be placed before `\maketitle'.
%%%Do NOT move this preamble from here!

\maketitle

\begin{abstract}
We study the magnetic interaction between a neutron star and its disk
 by solving the time-dependent relativistic force-free equations. At the
 initial state, we assume that the dipole magnetic field of the neutron
 star connects the neutron star and its equatorial disk, which deeply
 enters into the magnetosphere of the neutron star. Magnetic fields are
 assumed to be frozen to the star and the disk. The rotation of
 the neutron star and the disk is imposed as boundary conditions. We
 apply Harten-Lax-van Leer (HLL) method to simulate the
 evolution of the star-disk system. We carry out simulations for (1)
a disk inside the corotation radius, in which the disk rotates faster
 than the star, and (2) a disk outside the corotation radius, in which
 the neutron star rotates faster than the disk. Numerical results
 indicate that for both models, the magnetic field lines connecting the
 disk and the star inflate as they are twisted by the differential
 rotation between the disk and the star. When the twist angle exceeds
 $\pi$ radian, the magnetic field lines expand with speed close to the
 light speed. This mechanism can be the origin of relativistic outflows
 observed in binaries containing a neutron star.
\end{abstract}

\section{Introduction}

Magnetic fields play essential roles in various activities observed in
radio pulsars and in X-ray pulsars. In X-ray pulsars, an accretion disk
surrounding a neutron star interacts with the magnetic field of the
neutron star. In low mass X-ray binaries (LMXBs), angular momentum
supplied from the accretion disk accelerates the rotation of the neutron
star. These objects are sometimes observed as rejuvenated radio pulsars
(e.g., \cite{B1982}). Even in
isolated radio pulsars, recent observations by the Chandra X-ray
satellite (\cite{W2000}; \cite{H2002}; \cite{PTKS2003}) revealed the
existence of disk-like distribution of hot plasmas around the pulsar and
the existence of bipolar jets. Magnetic interaction between the neutron
star and its surrounding disk can be the origin of jet-like outflows.

The interaction between a magnetized star (neutron star, white dwarf or
a young stellar object) and its accretion disk has been studied
extensively. When magnetic field lines connect the star and its disk,
their torque spins up or down the central star (e.g., \cite{GL1978},
\yearcite{GL1979a}, \yearcite{GL1979b}; \cite{LRB1995}). 

Zylstra (1988) obtained a sequence of twisted force-free fields around a
neutron star and showed that when critical twist is accumulated, loss of
equilibrium leads to the expansion of the magnetosphere. Lynden-Bell and
Boily (1994) used semianalytic techniques for non-relativistic
force-free configurations. By studying the evolution of force-free
magnetic loops anchored to the star and the disk, they obtained
self-similar solutions for it. They found that twist injection from the
rotating disk makes the magnetic loops unstable and inflate and that the
loops expand along the direction $60^\circ$ from the rotational axis of
the disk. Lovelace \etal\ (\yearcite{LRB1995}) applied this mechanism to
keplerian disks.

In non-relativistic MHD case, Hayashi, Shibata, and Matsumoto (1996)
carried out two-dimensional (axisymmetric) magnetohydrodynamic
simulations of magnetic interaction between a star and its disk. They
showed that magnetic reconnection takes place in the current sheet
created inside the inflating magnetic loops. Their model can explain the
X-ray flares and outflows observed in protostars. Miller and Stone
(1997) presented the results of simulation including the rotation of the
central star. Goodson \etal\ (1997, 1999) and Goodson and Winglee (1999)
carried out longer time scale simulations of star-disk magnetic
interaction and demonstrated quasi-periodic formation and destruction of
magnetosphere by twist injection from the disk. Kato, Hayashi, and
Matsumoto (2004) reported the results of MHD simulation of the magnetic
interaction between a neutron star and its disk. They showed that due to
the gas pressure external to the expanding loop, magnetic tower
(\cite{L1996}) is formed and that semi-relativistic jet flows out along
the magnetic tower. This mechanism can explain the origin of
semi-relativistic jets observed in Sco X-1 (\cite{FGB2001a},
\yearcite{FGB2001b}). In their simulation, however, special relativistic
effects were neglected.

In magnetospheres of neutron stars, relativistic effects are not
negligible. The purpose of this paper is to extend the disk-star magnetic
interaction model to a relativistic regime. When plasma density is
low or magnetic field is strong, the Alfv\'{e}n speed can be close to
the light speed. Thus, we need to solve relativistic equations to study
the evolution of the magnetosphere. Here we present numerical results
for the magnetosphere dominated by the electromagnetic field, which can be
described by time dependent relativistic force-free equations. 

Recently, \citet{K2002} carried out simulations of relativistic
force-free fields to study the electrodynamics of pulsar magnetospheres and
black holes. Here we extend the model for the case including a rotating
equatorial disk threading the magnetosphere.

In section $2$, we describe the force-free equations, numerical scheme,
and redefinition of velocity fields. The results of simulations are
given in section $3$. Section $4$ is devoted for summary and discussion.  

\section{Numerical models}

\subsection{Basic equations}
We use force-free equations (e.g. \cite{uchi} for basic theory of
force-free fields) derived from the special relativistic
magnetohydrodynamics (SRMHD) equations to simulate the pulsar
magnetosphere. The force-free approximation is applicable when the
electromagnetic energy density is much larger than the energy density
of the plasma. The force-free equations are
\begin{equation}
\frac{\partial {\boldsymbol P}}{\partial t}+\nabla \cdot {\boldsymbol {\mathcal M}}=0, \label{eq:momentum}
\end{equation}
\begin{equation}
\frac{1}{c} \frac{\partial {\boldsymbol B}}{\partial t}+\nabla\times{\boldsymbol E}=0, \label{eq:induction}
\end{equation}
where
\begin{equation}
{\boldsymbol P}=\frac{1}{4\pi c}{\boldsymbol E}\times{\boldsymbol B} \label{eq:poynting}
\end{equation}
is momentum,
\begin{equation}
 {\mathcal M}^{ij}=-\frac{1}{4\pi}\left[ E^i E^j + B^i B^j - \frac{1}{2}\delta^{ij} \left( {\boldsymbol E}\cdot{\boldsymbol E}+{\boldsymbol B}\cdot{\boldsymbol B} \right) \right] \label{eq:tensor}
\end{equation}
is electromagnetic stress tensor and $c$ is speed of light. 
Equations (1), (2), (3) and (4) are equivalent to Maxwell's
equations and the force-free condition,
\begin{equation}
\rho_e {\boldsymbol E}+\frac{1}{c}{\boldsymbol J}\times{\boldsymbol B}=0,
\label{eq:ff}
\end{equation}
where $\rho_e$ is charge density and ${\boldsymbol J}$ is current
density. Equation (\ref{eq:ff}) implies degeneracy of electrmagnetic fields,
\begin{equation}
{\boldsymbol E}\cdot{\boldsymbol B}=0,
\label{eq:degeneracy}
\end{equation}
 which replaces the perfect conductivity
condition in classical MHD (see \cite{K2002}). Under this
condition, we can solve equation
(\ref{eq:poynting}) as
\begin{equation}
{\boldsymbol E}=-\left(\frac{B^2}{4\pi c}\right)^{-1} {\boldsymbol P} \times {\boldsymbol B}. \label{eq:electric0}
\end{equation}
We adopt ${\boldsymbol P}$ and ${\boldsymbol B}$ as fundamental
 variables for time dependent simulation. We normalize
the equations by using the typical strength of magnetic field $B_0$ (on the
surface of central star at rotational axis), and radius of the central star
$R_0$ as units. The unit of time is $\tau_0=R_0/c$. We use a spherical
coordinate system ($r$,$\theta$,$\phi$) with $\theta =0$ parallel to the
star's rotational axis. We assume axisymmetry ($\partial / \partial \phi=0$) 
but all three components of $\boldsymbol P$ and $\boldsymbol B$ are
retained. We assume that the star's dipole moment is aligned with the
rotational axis.

\subsection{Definition of velocity fields}

We introduce the velocity of magnetic field lines $\boldsymbol{v}_B$ as 
\begin{equation}
{\boldsymbol v}_B=\left( \frac{B^2}{4\pi c^2}\right)^{-1}{\boldsymbol P}. \label{eq:velocity}
\end{equation}
This determines the velocity of the magnetic field lines uniquely. 
By using (\ref{eq:velocity}), the electric field $\boldsymbol{E}$ is written as
\begin{equation}
\boldsymbol{E} = -\frac{1}{c}\boldsymbol{v}_B \times \boldsymbol{B}.
\label{eq:electric2}
\end{equation}
When the electric field is represented like equation (\ref{eq:electric2}), the
 electromagnetic field must satisfy
\begin{equation}
B^2 - E^2 \geq 0.
\label{eq:LI}
\end{equation}
We checked that the condition (\ref{eq:LI}) is satisfied throughout the
calculation. Since the force-free equations are derived from the
relativistic MHD equations by retaining the electromagnetic field only,
plasma's velocity $\boldsymbol{v}$ does not appear explicitly in the
theory. However, we can assume the existence of the plasma that is
frozen to the magnetic fields. If equation (\ref{eq:LI}) is satisfied,
the plasma can flow along the magnetic field lines.

\subsection{Numerical scheme}
 We use the HLL method (\cite{HLL1983};
\cite{jan}) to solve the time dependent equations. It belongs to the
family of an upwind scheme. In contrast to the Roe type scheme which
requires eigenvalues and eigenvectors of the characteristic matrix, HLL
scheme only needs eigenvalues. We used MUSCL-type method to attain
second order accuracy in space. The HLL method is formulated as follows
(\cite{HLL1983}). A one dimensional hyperbolic system can be written in
the form
\begin{equation}
\frac{\partial {\boldsymbol U}}{\partial t}+\frac{\partial {\boldsymbol F}}{\partial x}={\boldsymbol S}, \label{eq:hyperbolic}
\end{equation}
where the vector of conservative variables is
\begin{equation}
{\boldsymbol U}=(P^i,B^i)^t ,
\end{equation}
${\boldsymbol F}={\boldsymbol F}({\boldsymbol U})$ is the flux vector,
and ${\boldsymbol S}={\boldsymbol S}(x,t)$ is the source term. The
fluxes are defined at cell interfaces. We denote the interface variables
at the right- and left-hand side of each cell interface as 
${\boldsymbol U}_{\mathrm{R}}$ and ${\boldsymbol U}_{\mathrm{L}}$. We
denote the maximum speed of right- and left-going wave as
$b_{\mathrm{R}}$ and $b_{\mathrm{L}}$ (in this simulation,
$b_{\mathrm{R}}=+c$ and $b_{\mathrm{L}}=-c$) and the fluxes as 
${\boldsymbol F}_{\mathrm{R}}={\boldsymbol F}({\boldsymbol U}_{\mathrm{R}})$ 
and ${\boldsymbol F}_{\mathrm{L}}={\boldsymbol F}({\boldsymbol U}_{\mathrm{L}})$. 
The HLL flux is then given by
\begin{equation}
{\boldsymbol F}_{\mathrm{HLL}}=\frac{b_{\mathrm{L}}{\boldsymbol F}_{\mathrm{R}}+b_{\mathrm{R}}{\boldsymbol F}_{\mathrm{L}}-b_{\mathrm{R}} b_{\mathrm{L}}({\boldsymbol U}_{\mathrm{R}}-{\boldsymbol U}_{\mathrm{L}})}{b_{\mathrm{R}}-b_{\mathrm{L}}}. \label{eq:HLL}
\end{equation}
Second order accuracy in time is achieved by first advancing
${\boldsymbol U}$ by $\Delta t / 2$ and next advancing ${\boldsymbol U}$
by $\Delta t$ using the flux at $\Delta t / 2$.   

We use $450$ (model 1) and $460$ (model 2) non-uniform grids in the
radial direction and $180$ uniform grids in $\theta$-direction. The grid
size is $\Delta r=0.0125$ when $r<1.125R_0$ and gradually increases as
$\Delta r_{k+1} / \Delta r_{k}=1.01$  when $r>1.125R_0$. The maximum
radius is $r_{\mathrm max}=100R_0$ (model~1) and $r_{\mathrm
max}=110R_0$ (model~2). 

\subsection{Boundary conditions}
  We assume that dipole magnetic fields initially connect the central star and the
  geometrically thin disk at the equatorial plane ($3.0\leq r\leq  95.0$). 
 We assume that the central star and the disk are perfect conductors. At
  the equatorial plane, we impose boundary conditions such that $B_r$
  and $B_\phi$ are antisymmetric and $B_\theta$ is symmetric with
  respect to $\theta=\pi/2$. For electric fields, we impose the following
  conditions at $\theta=\pi/2$ ($3.0\leq r\leq  95.0$),
\begin{equation}
\left\{
\begin{array}{l}
 E_r=\displaystyle{\frac{r \omega}{c}}B_\theta, \\ 
 E_{\theta}=-\displaystyle{\frac{r \omega}{c}}B_r, \\
 E_\phi=0,
\end{array}
\right.
\label{eq:ED}
\end{equation}
which imply that magnetic fields rotate with the disk. Here $\omega$ is
the angular velocity of the Keplerian disk ($\omega=\sqrt{GM/r^3}$,
where $G$ and $M$ are gravitational constant and mass of the central star,
respectivery). At the equatorial plane in $1.0< r<3.0$, we imposed the
condition that $E_r$ and $E_\phi$ are symmetric and $E_\theta$ is
antisymmetric with respect to $\theta=\pi/2$. The magnetic field
component $B_{\theta}$ at the equatorial plane is extraporated from
$B_{\theta}$ at the grid point next to the equator. The electric field
$E_r$ at the equator is computed by using equation (\ref{eq:ED}). The
momentum ${\boldsymbol P}$ at the equator is computed by using equation
(\ref{eq:poynting}) and (\ref{eq:ED}).

The electric field on the stellar surface satisfies 
\begin{equation}
\left\{
\begin{array}{l}
E_r =\displaystyle{\frac{R_0 \Omega \sin{\theta}}{c}}B_{\theta}, \\
E_{\theta}=-\displaystyle{\frac{R_0 \Omega \sin{\theta}}{c}}B_r, \\
E_\phi =0,
\end{array}
\right.
\label{eq:ES}
\end{equation}
where $\Omega$ is the angular velocity of the central star. The magnetic
fields $B_r$, $B_{\theta}$ and $B_\phi$ at the stellar surface are extraporated from
$B_r$, $B_{\theta}$ and $B_\phi$ at the point next to the stellar surface. The
electric field at the stellar surface is computed by using equation
(\ref{eq:ES}). The momentum ${\boldsymbol P}$ is computed by using
equation (\ref{eq:poynting}) and (\ref{eq:ES}). In order to check that
the footpoints of magnetic field lines don't slip on
the surface of the star and the disk, we computed the velocity of
magnetic field lines near the surface of the star and the disk by using
equation (\ref{eq:velocity}). We confirmed that ${\boldsymbol v}_B$
coincides well with the rotational velocity of the
accretion disk and the stellar surface.

We carried out simulations for two models. In model 1, we take
$\Omega=0$ and $\omega =0.3 (c/R_0)(r/R_0)^{-3/2}$. In model 2, we take
$\Omega=0.05c/R_0$ and $\omega =0$. Model 2 corresponds to the disk far
outside the corotational radius. We use free boundary conditions at the
outer boundaries ($r=r_{max}$). On the rotational axis ($\theta=0$), we
substituted the values $P_r$ and $B_r$ at the grid point next to the rotational
axis.

\section{Numerical results}
\begin{figure}
 \begin{center}
   \FigureFile(170mm,60mm){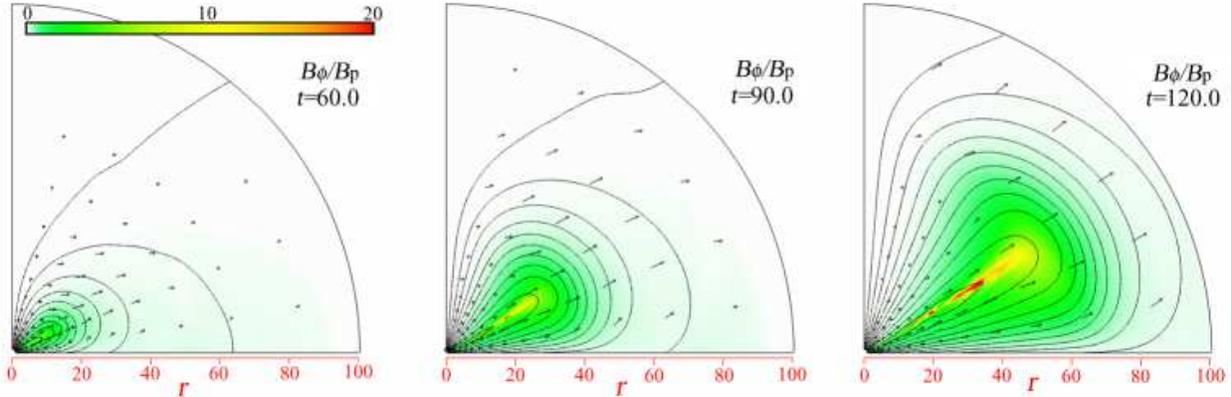}
 \end{center}
 \caption{Time evolution of magnetic field lines projected onto the
 $r-\theta$ plane (solid curves) in model 1. The color scale shows the
 twist $B_{\phi}/B_{\mathrm{p}}$. Arrows show the poloidal
 component of the velocity of magnetic field lines.}\label{Model1}
\end{figure}
Results of simulations for model 1 are shown in figure \ref{Model1}.
Solid curves in figure \ref{Model1} show the magnetic field lines
projected onto the $r$-$\theta$ plane. The color scale shows the
strength of twist of magnetic field ($B_{\phi} / B_{\mathrm{p}}$)
where $B_{\phi}$ is the toroidal magnetic field and
$B_{\mathrm{p}}$ is the poloidal magnetic field. The arrows show 
the poloidal component of the velocity of magnetic field lines defined by
equation (\ref{eq:velocity}). 

The disk at $r=3R_0$ rotates about $7$ radian at $t=120.0$. By the
rotation of the disk, the magnetic field lines connecting the central
star and the disk are twisted. Torsional Alfv\'{e}n waves propagating
along magnetic field lines are reflected on the surface of the central
star. The typical crossing time of Alfv\'{e}n waves travelling between
the central star and the disk is $\sim 2r/c$. As time goes on,
oscillations driven by the torsional Alfv\'{e}n wave dissipates.
 Since the toroidal component of
magnetic fields increases near the central star by accumulation of
twist, magnetic pressure increases. From about $t=50.0$, magnetic field lines
begin to inflate. This rapid expansion of twisted magnetic loops takes
place when the twist angle exceeds about $\pi$ radian. The inflation is
due to the loss of equilibrium in the twisted magnetic loops. Strong
current layer is created inside the expanding magnetic loops. These
results are consistent with the stability analysis by van Ballegooijen
(1994) and Lynden-Bell and Boily (1994).
\begin{figure}
 \begin{center}
   \FigureFile(170mm,60mm){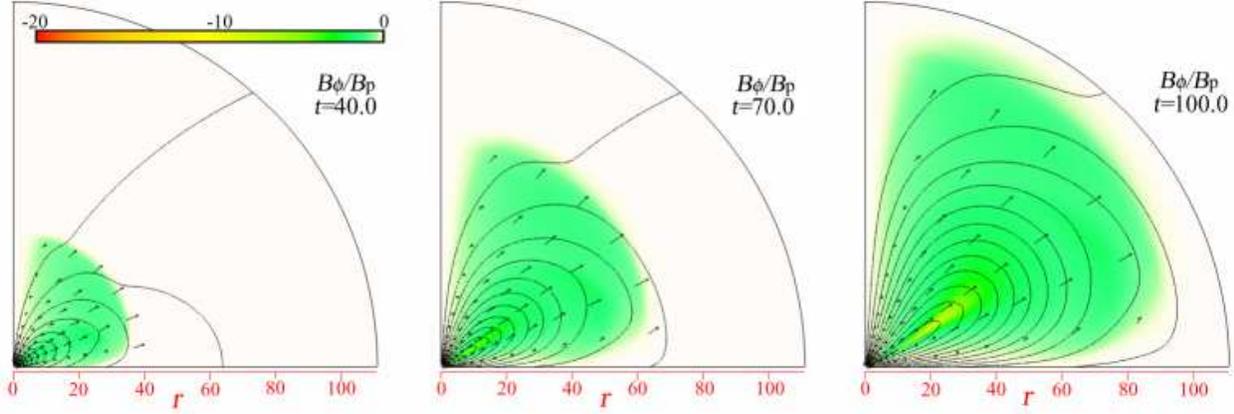}
 \end{center}
 \caption{The same as figure \ref{Model1} but for model 2. Solid curves
 show magnetic field lines. Color scale shows the twist
 $B_{\phi}/B_{\mathrm{p}}$. Arrows show the poloidal
 component of the velocity of magnetic field lines.}\label{Model2}
\end{figure}
Figure \ref{Model2} shows the time evolution of magnetic field lines
(solid curves), twist ($B_{\phi}/B_{\mathrm{p}}$), and
the poloidal component of the velocity of magnetic field lines (arrows)
for model 2. In this model, the central star is rotating faster than the
equatorial disk. Similarly to the model 1, the magnetic loops rapidly
expand after $t=50$ and form a long thin current sheet.
\begin{figure}
\begin{tabular}{cc}
\begin{minipage}{0.5\hsize}
 \begin{center}
   \FigureFile(75mm,40mm){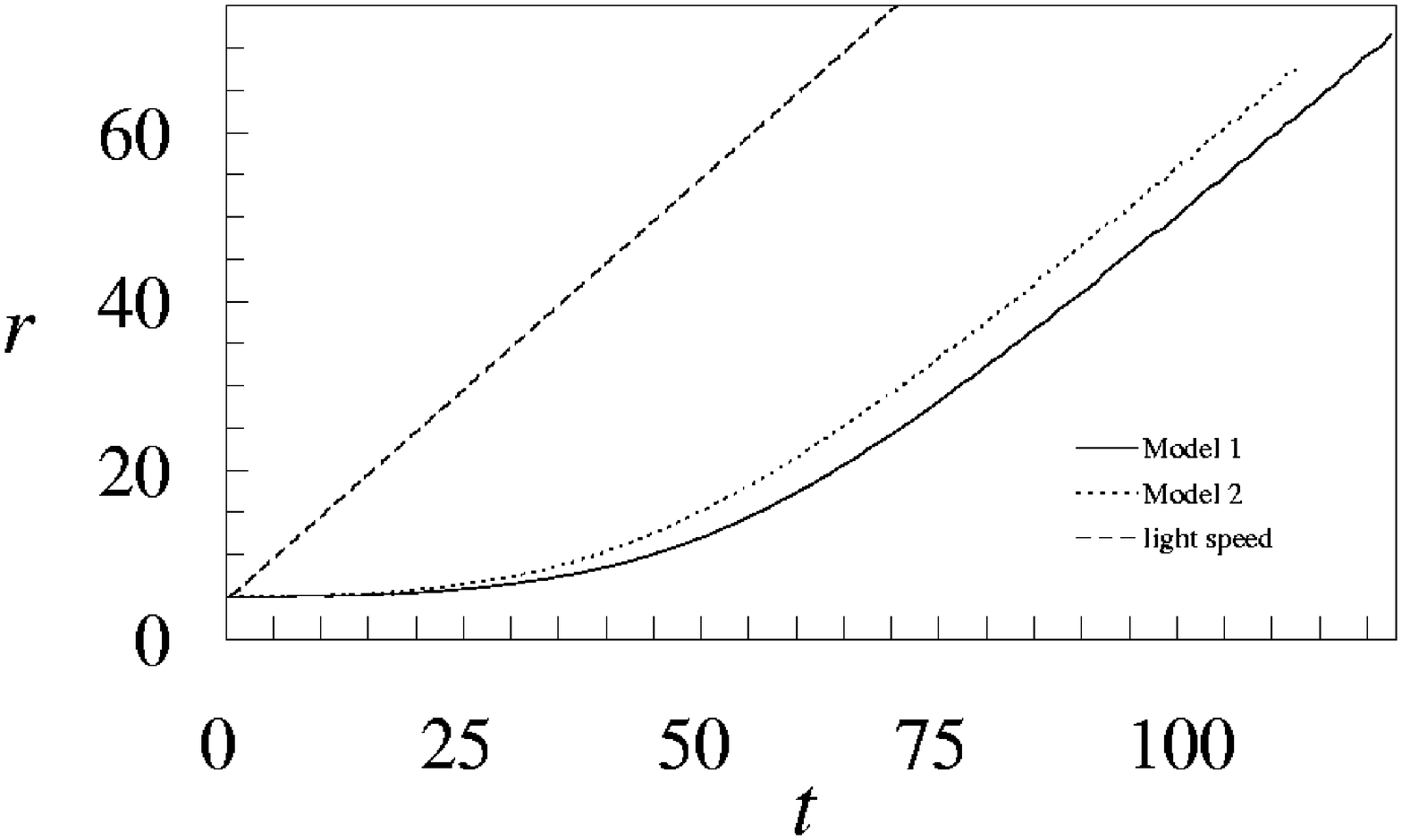}
 \end{center}
\end{minipage}
\begin{minipage}{0.5\hsize}
\begin{center}
   \FigureFile(85mm,50mm){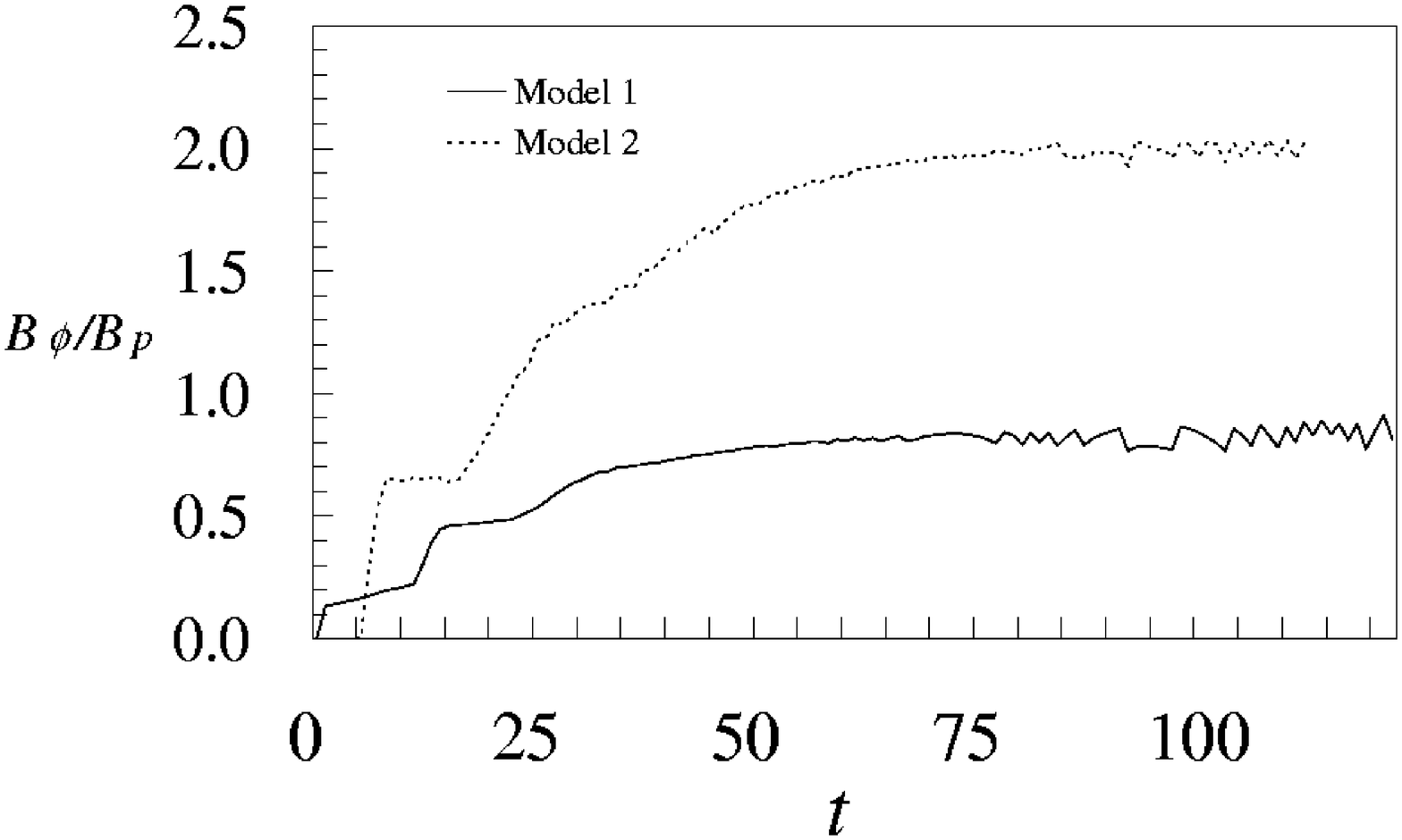}
\end{center}
\end{minipage}
\end{tabular}
 \caption{(Left) Time evolution of the maximum distance from $r=0$ to
 the point on the magnetic field line anchored to the disk at
 $r=4.9$. The solid curve is for model 1 and the dotted curve is for
 model 2. The dashed line represents the wave propagating at light speed.
 (Right) Time evolution of the twist $B_{\phi}/B_{\mathrm{p}}$ at
 the same point as that in the left figure. The solid curve is for model
 1 and the dotted curve is for model 2.}\label{data1}
\end{figure}
The solid and dotted curves in figure \ref{data1} (left) show the
maximum distance from $r=0$ to the magnetic field line anchored to the
equatorial disk at $r=4.9$. The expansion speed approaches the light
speed (dashed line). Figure \ref{data1} (right) shows
the degree of twist ($B_{\phi}/B_{\mathrm{p}}$) at the same point
as that of figure \ref{data1} (left). The twist increases with time until
$t \sim 50$. Subsequently, the magnetic field lines inflate almost
keeping the degree of twist.
\begin{figure}
\begin{tabular}{c}
\begin{minipage}{1.0\hsize}
 \begin{center}
   \FigureFile(80mm,50mm){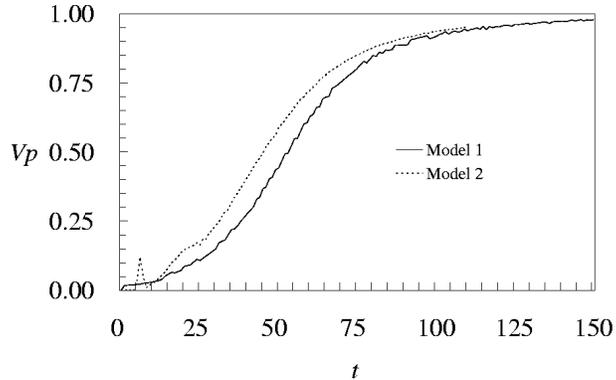}
 \end{center}
\end{minipage}
\end{tabular}
 \caption{ The poloidal velocity of the magnetic
 field line at the same point as those depicted in figure
 \ref{data1}. The velocity of magnetic field lines is defined by equation
 (\ref{eq:velocity}). The solid curve is for model 1 and the dotted
 curve is for model 2.}\label{data2}
\end{figure}
 Figure \ref{data2} shows the poloidal
 velocity of the magnetic field line at the same point as those
 depicted in figure \ref{data1}. The poloidal velocities in model 1 and 2
 exceed $90\%$ of the light speed at $t\sim 100$ and gradually approach
 the speed of light. Consequently, the toroidal velocity decreases and
 approach to zero.
\begin{figure}
\begin{tabular}{c}
\begin{minipage}{1.0\hsize}
 \begin{center}
   \FigureFile(80mm,50mm){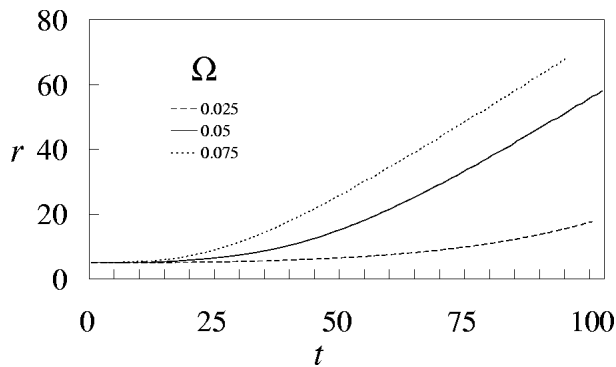}
 \end{center}
\end{minipage}
\end{tabular}
 \caption{The dependence of the radial expansion of magnetic fields on
 the angular speed of the central star, $\Omega$, when $\omega=0$ in the model 2. The curves show
 the maximum distance from $r=0$ to the point on the
 magnetic field line anchored to the disk at $r=4.9$. The solid curve is
 for model 2. The dotted
 curve is for $\Omega=0.075$, and the dashed curve is for $\Omega=0.025$.}\label{data3}
\end{figure}

 Figure \ref{data3} shows the dependence of the radial expansion of
 magnetic fields on the angular speed of the central star, $\Omega$, when
 $\omega=0$. The solid curve is for model 2. As $\Omega$ increases, the
 poloidal component of the velocity of magnetic field lines approaches
 the light speed within shorter period.

\section{Summary and Discussion}

In this paper, we numerically solved the relativistic force-free
degenerate electrodynamic equations to study the time evolution of
magnetic field lines connecting a neutron star and its disk. This
approach is an extension of the study by \citet{UKL2002} in which they
solved force-free magnetosphere by using semianalytic methods. Starting
from potential magnetic fields, we showed that magnetic fields are
twisted by the differential rotation between the disk and the central
star. After critical twist accumulates, the magnetic field lines
connecting the star and the disk inflate with speed exceeding
$0.95c$. In the region distant from the central star and the disk, the
poloidal velocity approaches the speed of light. When the plasma is loaded
to these expanding magnetic loops, we can expect relativistic outflows
with Lorentz factor $\Gamma \ge 3$. A long thin current sheet is created
inside these expanding magnetic loops. The elongation of the current
sheet continues until the end of the calculation.

In this paper, we have solved ideal force-free equations. When the
resistivity is included, magnetic reconnection will occur in the current
sheet. Since force-free equations can't handle magnetic reconnection, we
need to solve the full set of resistive MHD equations to simulate
the magnetic reconnection. Another limitation of force-free approximation is
that gas pressure does not appear in force-free equations. The existence
of plasma pressure or magnetic fields helps collimating the expanding
magnetic field lines toward the rotational axis (e.g. \cite{KHM2004};
\cite{TB2002}). Relativistic MHD simulations including gas pressure will
be able to reproduce the collimated relativistic jets. We will report
the results of such simulations in subsequent papers. 
\begin{verbatim}
\end{verbatim}

 The authors thank Y. Uchida, S. Hirose, M. Nakamura for discussion. We
 thank S. Komissarov for discussion during the visit of R. Matsumoto to
 the University of Leeds. Numerical computations were carried out on VPP5000 at
 the Astronomical Data Analysis Center, ADAC, of the National
 Astronomical Observatory, Japan. This work is supported by the grant of
 ministry of education, science, sports, culture and technology
 (15037202, P.I. R.~Matsumoto) and the Japan Society for the Promotion of Science
 Japan-UK Cooperation Science Program (P.I. K. Shibata and N. O. Weiss).

\end{document}